\documentstyle[12pt]{article}
\begin{document}
\pagestyle{empty}
\begin{center}  \LARGE MOLECULAR LINES IN BOK GLOBULES \end{center}
\begin{center}   \LARGE AND AROUND HERBIG Ae/Be STARS   \\
\vspace*{1.5cm}
\Large  F. Scappini$^1$, G.G.C. Palumbo$^2$, G. Bruni$^1$, and
\vspace{1.5cm} P. Bergman$^3$ \\
\large
$^1${\it Istituto di Spettroscopia Molecolare, C.N.R.,
 Via de' Castagnoli, 1 -  40126 Bologna, Italy} \\ \vspace*{0.5cm}
$^2${\it Dipartimento di Astronomia, Universit\`a degli Studi,
    Via Zamboni, 33 and I.T.E.S.R.E./C.N.R.
Via de' Castagnoli, 1 - 40126 Bologna, Italy
} \\  \vspace*{0.5cm}
$^3${\it Onsala Space Observatory, S - 43900 Onsala, Sweden}
\end{center}
\newpage
\begin{center} \section*{ABSTRACT} \end{center}
\hspace*{0.7cm} This paper is intended as part of a more extensive
molecular line
survey in star forming regions along the evolutionary track of a
collapsing cloud toward a young stellar object. \\
\hspace*{0.7cm} We have studied a sample of seven small dark clouds
(Bok globules) and eight Herbig Ae/Be stars in the J=1$\rightarrow$0
transition of HCO$^{+}$, H$^{13}$CO$^{+}$, HCN and H$^{13}$CN.
The choice of these molecules is determined by the simple chemistry
and the predicted high abundance of the reactants leading to their
formation. The isotopically substituted species (isotopomers),
H$^{13}$CO$^{+}$ and H$^{13}$CN, were observed in order to determine,
whenever possible, the optical thickness of the main species.
The most abundant isotopomers were found in almost all the sources
(detection rate 70-90\%). Those sources which exhibited the strongest
signals were also searched for the $^{13}$C isotopomers. H$^{13}$CO$^{+}$
was found in one dark cloud and around three Herbig Ae/Be stars,
while H$^{13}$CN around only one star. The column densities
for each species and the physical conditions of the objects were derived
whenever the observational data allowed it.
\newpage   \noindent
Please send proofs to: \\  \\
\noindent
\hspace*{3cm} dr. Flavio Scappini \\
\hspace*{3cm} Istituto di Spettroscopia Molecolare, CNR \\
\hspace*{3cm} Via de' Castagnoli, 1 \\
\hspace*{3cm} 40126 Bologna, Italy \\
\newpage
\pagestyle{plain}
\normalsize
\begin{center} \section*{I. INTRODUCTION} \end{center}
\hspace*{0.7cm} Molecular line observations are the most
versatile probes of the
internal properties of star forming regions. Molecular transitions
can provide close investigations of the mechanisms which govern star
formation processes. Thus, the composition, morphology and
kinematics of the high density gas, which are relevant to
the understanding of the phenomena associated with star formation,
can be studied through observations of molecular tracers
(Hartquist {\it et al.}, 1993).
The physical processes that accompany the birth and early evolution of a
star modify the chemistry in the surrounding medium through increases in
density and temperature and through very energetic mass ejection.
Bipolar outflows of material
in the neighbourhood of young stars, shock fronts with associated
Herbig-Haro objects, maser emissions (H$_2$O, OH, SiO, CH$_3$OH,
etc.) and highly excited rotational molecular lines of many species
are only a few of the typical
manifestations of newly-born stars interacting with their parental
molecular clouds (Lada, 1985).

In order to contribute data to the understanding of the star forming
processes we have measured HCO$^+$ and HCN millimeter-wave emissions
in two different sample regions: (i) Bok globules and (ii) Herbig Ae/Be
stars.

Bok globules, also called Barnard objects, are nearby (150-600 pc)
dense interstellar clouds ($n(H_{2}$)=10$^3$-10$^5$ cm$^{-3}$) of
gas and dust.
In this sense they have to be considered a subsample of small dark clouds.
They appear visibly as well defined dark patches with typical sizes
of $\sim$ 1 pc against the general background of stars. While most
globules appear to be gravitationally bound and in virial equilibrium,
recent infrared and radio observations suggest that low-mass (1 -2 M$\odot$)
star formation is taking place in some of them (Clemens and Barvainis, 1988,
hereafter CB; Yun and Clemens, 1992, hereafter YC).

The Herbig Ae/Be stars are objects of intermediate mass ($\sim$ 3 M$\odot$)
in their pre-main
sequence stage of evolution. Some of these stars show evidence of mass
ejection. Because of their youth they are still embedded in the dark
clouds which have originated them (Herbig, 1960; Finkenzeller and Mundt,
1984).

These two classes of objects represent a wide scenario of situations.
Bok globules range from quiescent to star forming clouds and
Herbig Ae/Be stars are already pre-main sequence objects still in some
dynamical interactions with the environment.
The study of molecular
spectra from these regions should allow us to characterize the chemical and
physical conditions of the different stages along the evolution sequence
of a small molecular cloud toward star formation.

In this paper we report results of selected molecular line
observations in seven Bok globules and in the cloud material
surrounding eight Herbig Ae/Be stars. The number of objects and lines
have been limited by telescope time allocation and frequency bands
availability.
We searched for the
J=1$\rightarrow$0 rotational transitions of HCO$^+$, H$^{13}$CO$^+$,
HCN, and H$^{13}$CN using the 20m radiotelescope at Onsala. These molecules
have been chosen in view of the simple chemistry that leads to their
formation and to the predicted abundance of the precursor species,
CO, H$_{3}^{+}$, C$^{+}$, N, etc.(Hartquist {\it et al.}, 1993). The
isotopically substituted species, H$^{13}$CO$^{+}$ and H$^{13}$CN,  have been
searched for in order to determine, if detected, the optical thickness of the
main species. In paragraph II observations at the telescope are described.
In paragraph III the criteria adopted to select the two observed samples are
illustrated; for each source a brief description is given. Results are
summarized in paragraph IV and the source physical properties derived.
A summary of the paper is given as conclusion.

\begin{center} \section*{II. OBSERVATIONS} \end{center}
\hspace*{0.7cm}
Observations were made in February and May 1990 with the 20m
radiotelescope at Onsala. The half-power beamwidth of the
telescope was 45" at 90 GHz. The spectrometer used was a 256-channel
filter-bank with 250 kHz resolution (0.83 kms$^{-1}$).
The SSB-tuned SIS receiver was operated in a dual beam switching
mode and a chopper wheel calibration technique was used. The
main beam efficiency ($\eta_{mb}$) was determined to be 0.43. We
position switched, with one on-source per off-source measurement and
90s integration time on each position. The pointing error, checked
every three hours, was found to be better than 4" rms on all occasions.

\begin{center} \section*{III. DESCRIPTION OF THE SAMPLE} \end{center}
\hspace*{0.7cm} In  Table 1 we present a tabulation of the sources,
consisting of seven Bok globules and
eight Herbig Ae/Be stars. The source name is
followed by the location of the central position in 1950 equatorial
coordinates. The presence of water maser
emission is also indicated, whenever detected.

Bok globules are dense condensations and possible sites of
star formation (CB). They have optical sizes
between 1' and 10'. Frerking and Langer (1982) and Wootten
{\it et al.} (1982) detected in them different molecular species.
Recent IRAS observations have confirmed that many of these dark clouds
have internal heat sources (Beichman, 1986), some of which have
infrared colours corresponding to T-Tauri stars. For the choice of the
Bok globules to be observed we adopted the following criteria:
(i) to be IRAS point sources,
(ii) these point sources have infrared colours corresponding
 to those of Beichman (1986) and Emerson (1987) samples
for T-Tauri stars and cores. The selected objects in a colour/colour
diagram show the following distribution:
-0.86 $\leq$ log (F$_{12}$/F$_{25}$) $\leq$ -0.17, -1.44 $\leq$ log
(F$_{25}$/F$_{60}$) $\leq$ -0.29, and -0.55 log (F$_{60}$/F$_{100}$) $\leq$
-0.03 (F$_{12}$, F$_{25}$, F$_{60}$, and F$_{100}$ are the infrared
fluxes at 12,25,60, and 100 $\mu$m, respectively).
Finally,  (iii) to have quite large CO linewidths. The above selection
criteria clearly privilege active star forming clouds.

Ae/Be stars were originally selected by Herbig (1960) as Ae or Be stars
embedded in nebulosities and later found to be young stellar
objects in the pre-main sequence phase (Finkenzeller and Mundt, 1984).
Our selection criteria are:
(i) they have optical outflows and (ii) P-Cygni profiles
(H$\alpha$ lines).

In the following a brief description of each source is given together with
its present observational situation. \\

\noindent
\underline{\large LBN594}.
Optical dimensions 6.7' $\times$ 5.6' and IRAS source
00259+5625.
It has been observed in CO (CB). Recently Scappini {\it et al.}, (1991)
have detected maser emission 1~arcmin south of the cloud center. CO outflow
has been discovered by YC. \\

\noindent
\underline{\large LBN613}.
Optical dimensions 9' $\times$ 3.4' and IRAS source 00465+5028.
It has been observed in CO (CB).          \\

\noindent
\underline{\large L1534}.
Also called TMC-1A, dark cloud in the Taurus cloud complex,
IRAS source
04365+2535. The area, 0.87 square degrees, is larger than that of the other
dark clouds of our sample (Lynds, 1962). It has been observed in CO by
Heyer {\it et al.} (1987) and in NH$_{3}$ by Benson and Myers (1989). \\

\noindent
\underline{\large CB34}.
Optical dimensions 4.5' $\times$ 2.2' and IRAS source
05440+2059 (CB). It showed CO outflow (YC). \\

\noindent
\underline{\large L810}.
Optical dimensions 10.1' $\times$ 6.7'and IRAS
source 19433+2743 (CB).
It has been observed in HCO$^+$, in NH$_3$ and in H$_2$CO by
Wootten {\it et al.}, 1982.
A near infrared source about at the center of the cloud has been
discovered by Nekel {\it et al.}, (1985), together with a nearby
6$_{16}$$\rightarrow$5$_{23}$
water maser emission.
More recently Harju (1989) has mapped the dark cloud in HCN(J=1$\rightarrow$0)
and measured the ratio of the integrated intensities HNC/HCN towards the cloud
center to be 0.4$\pm$0.1. CO outflow has been first discovered by
Xie and Goldsmith (1990).   \\

\noindent
\underline{\large L797}.
Optical dimensions 6.7' $\times$ 4.5' and
IRAS source 20037+2317 (CB). CO outflow has been discovered (YC).       \\

\noindent
\underline{\large L1262}.
Optical dimensions 11.2' $\times$ 5.6' and IRAS
source 23238+7401 (CB).
First observed in CO by Parker {\it et al.} (1988), showing evidence
of outflow associated with the embedded IRAS source.
A millimeter-wave interferometry map in CO by Tereby {\it et al.} (1989) has
confirmed the presence of a spatially compact low-velocity outflow.
The $^{12}$C/$^{13}$C abundance ratio was obtained by measurements of
the optically thin species $^{12}$C$^{18}$O and $^{13}$C$^{18}$O to be
75$\pm$8(1$\sigma$), close to the solar value 89. Similarly, the
$^{18}$O/$^{17}$O and $^{16}$O/$^{18}$O ratios were measured. \\

\noindent
\underline{\large LkH$\alpha$198}.
Ae star embedded in an anonymous dark
cloud in Cassiopea. It exhibits molecular outflow (Cant\`o {\it et al.}
1984; Levreault 1988). The outflow has been found to be quite large
extending asymmetrically to about 1.6 pc from the star with a total
extent of 2 pc. As pointed out by Ho {\it et al.} (1982)
an asymmetric bipolar outflow such as this can easily be explained
by the presence of density gradients in the ambient cloud.
The outflow mass is 3.7 M$_{\odot}$  (Levreault, 1988) and its velocity
6.7 km/s for the blue and 5.9 km/s for the red emission.
A number of molecular line observations were carried out,
besides those in CO and $^{13}$CO, by different authors
(Loren 1977; Loren 1981; Kogure 1988; Levreault 1988). Data
concerning the column density of HCO$^+$ and HCN were not
specifically reported, even if these molecules were detected. \\

\noindent
\underline{\large RRTau}.
A spectral type classification as A6 was given to this star by
Cohen and Kuhi (1979). A 3'$\times$ 3' map in CO shows no outflow
(Cant\`o {\it et al.}, 1984). \\

\noindent
\underline{\large HD250550}.
The star is classified as B6 by Cohen and
Kuhi (1979). P-Cygni profiles in the Balmer lines were observed by
Herbig (1960) and more recent CO mapping (Cant\`o {\it et al.} 1984)
showed evidence of peculiar activity in the direction of the NE
arc-shaped associated nebulosity. \\

\noindent
\underline{\large BD463471}.
The H$\alpha$ lines have P-Cygni
profiles with absorption at about -200 Km/s (Herbig, 1960).
A bipolar outflow was observed (Levreault, 1988) as well as infrared
emission in the 1.5 - 2.3 $\mu$m range (Harvey, 1984). The interpretation
of this emission is that it arises from recombination in a HII
region around the star. \\

\noindent
\underline{\large V645Cyg}.
Known from radio-observations to be associated with high
velocity molecular gas (Bally and Lada, 1983). The
6$_{16}$$\rightarrow$5$_{23}$
water maser emission was, in fact, found at -46.1 kms$^{-1}$
(Lada {\it et al.}, 1981). \\

\noindent
\underline{\large LkH$\alpha$234}.
Ae star contained in the bright nebulosity
NGC 7129 together with the Be star BD 651637 (Loren, 1977). Both
stars were classified by Herbig (1960) as young massive stars. However
the star BD 651637 is in a more advanced stage of evolution and has
had more time to disperse the circumstellar gas (Strom {\it et al.}, 1972).
It has been shown that the maximum of the T$_{A}^{*}$(CO)
in the NGC 7129 cloud is associated with LkH$\alpha$234
rather than with BD 651637 (Loren, 1977). The 6$_{16}$$\rightarrow$5$_{23}$
water maser emission was observed at -14.9 kms$^{-1}$ (Rodriguez {\it
et al.}, 1980). Besides having been mapped in CO, other species,
such as CS, SO, HCN, H$_2$CO, and HCO$^+$ were detected in the
surrounding molecular cloud. However the somewhat limited angular
resolution (2.6' beamwidth) prevented detailed information on
the molecular distribution (Loren {\it et al.}, 1977). \\

\noindent
\underline{\large LkHa233}.
The spectral type is estimated as A7 (Herbig, 1960).
Carbon monoxide emission in both the CO and $^{13}$CO
J=1$\rightarrow$0 lines has been observed from the region
surrounding the star (Loren {\it et al.} 1973). From H$_2$CO
observations the density of the cloud core was estimated to be
about 5x10$^3$ cm$^{-3}$ and the kinetic temperature 28K
(Loren, 1981). The star excites the NGC 1788 nebula, which is embedded
in a small dust cloud (Loren, 1981). \\

\noindent
\underline{\large MWC1080}.
Classified as B0 by Cohen and Kuhi (1979). The CO outflow map
shows that the regions of blushifted and redshifted emission are
concentric and centered on the star. This indicates either an
isotropic outflow or a bipolar outflow observed along its axis
(Cant\`o {\it et al.}, 1984; Levreault, 1988). A more detailed map
by Yoshida {\it et al.} (1991) suggests that the outflow is not exactly
isotropic but bipolar.
Herbig-Haro emission east of the star and poorly collimated has been found
by Poetzel {\it et al.} (1992). HCO$^+$ (J=1$\rightarrow$0) has been
observed by Koo (1989).

\begin{center} \section*{IV. RESULTS}   \end{center}
\begin{center} {\it a) The Survey Data} \\  \end{center}
\hspace*{0.7cm}
The transition frequencies of the J=1$\rightarrow$0
lines of HCO$^+$, H$^{13}$CO$^+$, HCN and H$^{13}$CN, together with the
electric dipole moments and the N-nuclear quadrupole coupling constants
are to be found in De Lucia and Gordy (1969), Pearson {\it et al.} (1976)
and Bogey {\it et al.} (1981).

It is worth summarizing the dominant reactions which produce the observed
species. For HCO$^+$,
\begin{eqnarray}
H_{3}^{+} + CO & \rightarrow & HCO^{+} + H_{2} \\
N_{2}H^{+} + CO & \rightarrow & HCO^{+} + N_{2} \\
CH_{3}^{+} + O & \rightarrow & HCO^{+} + H_{2}
\end{eqnarray}
Among these the most important is the reaction with the abundant species
$H_{3}^{+}$ (Huntress and Anicich, 1976). For HCN,
\begin{eqnarray}
CH^{+}_{3} + N & \rightarrow & H_{2}CN^{+} + H \\
NH_{3} + C^{+} & \rightarrow & H_{2}CN^{+} + H \\
HNC + HCO^{+} & \rightarrow & H_{2}CN^{+} + CO \\
HNC + H_{3}^{+} & \rightarrow & H_{2}CN^{+} + H_{2}
\end{eqnarray}
\noindent
$H_{2}CN^{+}$ which has three isomeric forms, leads by
dissociative recombination to HCN and HNC in a ratio which seems to
depend very much upon the ambient conditions (Walmsley, 1993).

Table 2 gives the results of the observations, that is the peak antenna
temperature T$_{A}^{*}$, corrected for atmospheric losses, the FWHM
linewidth, and the LSR velocity of the peak emission. A Gaussian
profile was numerically fitted to each line to obtain the peak antenna
temperature T$_{A}^{*}$, linewidth $\Delta$$v$  (FWHM), and center frequency.
In the case of HCN and H$^{13}$CN the nitrogen quadrupolar nucleus
splits the J=1$\rightarrow$0 line into three components corresponding to
F=1-1, F=2-1, and F=0-1, with theoretical intensity ratios 33:56:11,
respectively. These spectra were fitted simultaneously to the expected
hyperfine pattern of three lines. The spectra of MWC1080 (J=1$\rightarrow$0,
HCO$^{+}$), LBN594 (J=1$\rightarrow$0, HCN) and LkH$\alpha$233
(J=1$\rightarrow$0, HCN)
appear  contaminated with extra features and/or noise. They were fitted
using the above described procedure together with the known {\it v}$_{LSR}$
of CO emission (see Table 1).

The quoted T$_{A}^{*}$ for HCN and H$^{13}$CN in Table 2 is the sum
of the peak temperatures of the three components. All errors in
Table 2 are only 1$\sigma$ fitting errors and all limits are 3$\sigma$.

HCO$^+$ was detected in all seven Bok globules and around
six Herbig Ae/Be stars.
HCN was found in five globules and around five stars. In those sources
where HCO$^+$ and HCN exhibited the strongest signals we searched also
for H$^{13}$CO$^+$ and H$^{13}$CN. H$^{13}$CO$^+$ was found in
one Bok globule and around three Herbig Ae/Be stars and H$^{13}$CN
around one star.
Typical spectra are presented in Figs. 1 to 10.

{}From the spectra of HCN and H$^{13}$CN it can be seen that the
intensity ratios between the hyperfine components do not correspond
to the theoretical ones. In fact, while the two strongest lines retain
approximately their theoretical ratio, the third is more
intense than it should. Cernicharo {\it et al.}, 1984 have suggested
that in cold dark clouds scattering in the surrounding envelope
changes the HCN hyperfine ratios formed in the core so that the
optically thinnest line, F=0$\rightarrow$1, shows up enhanced
relative to the other two hyperfine components.

In Table 3 we present the integrated line intensities \( I(X) =
\frac{1}{\eta_{mb}} \int T_{A}^{*} dv \)
for
the observed species (X = HCO$^+$, H$^{13}$CO$^+$, HCN, and H$^{13}$CN).
The ratios I(HCO$^+$)/I(H$^{13}$CO$^+$) and I(HCN)/I(H$^{13}$CN)
are also given, whenever the corresponding intensities were measured.

HCO$^+$ and HCN have been mapped (9-10 points in steps of one
beam size) in CB34 and in
LkH$\alpha$234 and Figs. 11 and 12 show the corresponding integrated
line intensity distribution.
The position of the peak of the molecular emission coincides with the
position of the center of the cloud or of the star within, at most,
one beamwidth.

\begin{center} {\it b) Physical properties}  \end{center}
\hspace*{0.7cm}
We derive estimates of the optical depth from the ratio of the
$^{12}$C and $^{13}$C isotopomer line intensities, I($^{12}$C)/I($^{13}$C).
Assuming that the lines of the isotopomers are at the same
excitation temperature and that the beam filling factors are similar
(here they are taken to be unity) one has,
\begin{equation}
\frac{I(^{12}C)}{I(^{13}C)}=\frac{1-exp(-\tau_{12})}
{1-exp(-\tau_{13})}
\end{equation}
\noindent
where $\tau_{12}$ and $\tau_{13}$ are the optical depths of the
$^{12}$C and $^{13}$C isotopomers, respectively. Adopting the solar
isotopic abundance ratio $\tau_{12}$/$\tau_{13}$ = 89, this expression
can be solved for $\tau_{13}$. Table 4 reports the calculated
optical depths for HCO$^+$ and HCN, whenever the ratio
I($^{12}$C)/I($^{13}$C) has been determined.

The excitation temperature $T_{ex}$ can be calculated from the
observed value of T$_{A}^{*}$
in the optically thick case,
\begin{equation}
T_{ex} = ( h \nu_{10}/k ) \left\{ ln \left[ 1+ \frac{h \nu_{10}/k}
{J_{\nu}(T_{bg}) + T_{A}^{*}/ \eta_{mb}} \right] \right\} ^{-1}
\end{equation}
\noindent
where $h\nu_{10}/h$ $\cong$ 4 K at the frequency $\nu_{10}$ of the
J=1$\rightarrow$0 transition, the background temperature $T_{bg}$ = 2.7 K
and the function $J_{\nu}$ at background temperature is,
\begin{equation}
J_{\nu}(T_{bg})=\frac{h\nu_{10}/k}{exp(h\nu_{10}/_{k}T_{bg})-1}
\end{equation}
\noindent
Values of $T_{ex}$ derived in this manner,
for the ascertained optically thick cases, are listed in Table 4.
For the excitation temperature to be equal to the gas kinetic temperature
$T_{k}$ the critical density is calculated to be $n(H_2)\simeq 10^{6}$
cm$^{-3}$ and this condition may not be always fulfilled for the objects
under investigation.

The column densities, for optically thin emission (\( \tau<<1 \)),
are obtained with the relation,
\begin{equation}
N=\frac{8 \pi k^{2} \nu_{10}}{3h^{2}c^{2}A_{10}B_{0}} \mbox{    }
\frac{T_{ex}exp\left(h \nu_{10}/kT_{ex}\right)}
{1-(J_{\nu}(T_{bg})/J_{\nu}(T_{ex}))} \mbox{    }
\frac{1}{\eta_{mb}}\int T_{A}^{*}d\nu
\end{equation}
\noindent
where
A$_{10}$ is the Einstein coefficient for spontaneous emission,
B$_{0}$ is the rotational constant, and \( J_{\nu}(T_{ex}) \) is the
radiation temperature of a black body at temperature T$_{ex}$.

In the optically thick limit ($\tau>>1$) the integrated line
intensity has to be multiplied by the optical depth to obtain
corrected column densities.

Table 5 reports the column densities for the observed species calculated
using Eq.(11) in the optically thin limit and, assuming \( T_{ex} = 10K \)
for all sources (Myers, 1985). Only for those objects for which
the optical depth was calculated, see Table 6, an optical depth correction
factor was introduced.
The column densities of HCO$^+$ and HCN calculated in the remaining
objects may be underestimated. Moreover, the isotopomers H$^{13}$CO$^+$
and H$^{13}$CN are assumed to be optically thin in all the observed objects.

The assumed excitation temperature \( T_{ex}=10K \) in the Bok globules and
in the gas around
the Herbig Ae/Be stars is supported by the
values obtained in Table 4. From this temperature a thermal linewidth,
for the molecular observations reported here, is calculated to be
\( \Delta v \simeq 0.13\) kms$^{-1}$. The observations reported in
Table 2 show linewidths much larger than thermal $\sim$ 1.1-3.1 kms$^{-1}$,
implying bulk motions over some unknown length scale.

\begin{center} \section*{SUMMARY} \end{center}
\hspace*{0.7cm} In order to contribute molecular data on star forming
regions we have measured several molecular transitions in Bok globules
and Herbig Ae/Be stars. The selection criteria for the Bok globules
were intended to include those with characteristic features of possible
star formation. For the Herbig Ae/Be stars the criteria aimed at
objects with outflows.
Specifically, we have searched for the J = 1$\rightarrow$0 ground state
transition of HCO$^+$, H$^{13}$CO$^+$, HCN, H$^{13}$CN in seven
Bok globules and in eight Herbig Ae/Be stars. The detection rate was
very high for the normal isotopes (HCO$^+$ = 90\% and HCN = 70\%), and lower
for the $^{13}$C isotopomers, but these have not been searched in all
objects.

The HCO$^+$/HCN column density ratio distribution among the
investigated Bok globules is 0.3 - 1.3 and among the Herbig Ae/Be
stars is 0.4 - 1.9. This shows that there is a large overlapping
density interval between the two categories of objects. Table 6 compares
the HCO$^+$/HCN column density ratio, excitation temperature, density and
linewidth of the present sample with those of other known regions.
The chemical and physical scenario of the investigated objects looks similar
to that found in cold regions, but linewidths are larger than thermal
broadening alone. It is worth noting that in two Bok globules (LBN594
and L810) as well as in two Herbig stars (V645Cyg and LkH$\alpha$234)
water emission was found (Scappini {\it et al.}, 1991; Nekel {\it
et al.}, 1985; Lada {\it et al.}, 1981; Rodriguez {\it et al.}, 1980).

A few maps of the objects exhibiting the highest molecular abundance
give an idea of the total line area and of the almost symmetric
distribution around the object position.

Even if our results have to be considered very preliminary, still their
analysis suggests a quite similar scenario of abundances and physical
conditions in the observed Bok globules and in the gas around the
Herbig stars. These similarities together with the already discussed
features, such as embedded infrared sources, water maser emission, CO outflow,
and large CO linewidths, suggest that the presently studied
globules are likely to be sites of low-mass star formation.

As discussed in Section III, HCN is produced by dissociative
recombination of H$_{2}$CN$^{+}$ with electrons, and this produces
HNC as well. It would be interesting to search  for HNC and the
correlated species NH$_{3}$ in the same objects. In an investigation
over twenty dark clouds cores Harju (1989) has found that the average
HNC/HCN(J=1$\rightarrow$0) intensity ratio is of the order of one.
Furthermore the HNC distribution in L1551 is very similar to that
of ammonia, while HCN behaves somewhat differently, supporting the
idea of different chemical origin between the two isomers.

Future observations will be aimed at other molecular species and
also at more quiescent globules in order to eventually detect
composition differences between active and inactive regions.

\newpage
\section*{ACKNOWLEDGEMENTS}
\hspace*{0.7cm} We would like to thank OSO for technical support and
Prof. A. Dalgarno and Dr. L. Avery for critical reading of the manuscript.
F.S. and G.G.C.P. acknowledge financial support from Agenzia Spaziale
Italiana; G.G.C.P. was also supported by Ministero per l'Universit\`a e la
Ricerca Scientifica e Tecnologica.
We are also grateful
to Mrs. M.G. Balestri for data bank search and typing. A quite long
interaction with the referee has improved the paper to the present stage.
\newpage
\begin{center} \section*{References}   \end{center}
\begin{description}
\item{}Adams, F.C., and Lizano, S. 1987, Ann. Rev. Astr. Ap. 25, 23.
\item{}Bally, J. and Lada, C.J. 1983,     Ap. J. 265, 824.
\item{}Beichman, C.A., Myers. P.C., Emerson, J.P., Harris, S.,
Mathieu, R., Benson, P.J., and Jennings, R.E. 1986,  Ap. J.
307, 337.
\item{}Benson, P.J., and Myers, P.C. 1989,     Ap. J. Suppl. 71, 89.
\item{}Bogey, M., Demuynck, C., and Destombes, J.L. 1981     Mol. Phys. 43,
1043.
\item{}Cant\`o, J., Rodriguez, L.F., Calvet, N., and Levreault, R.M. 1984,
    Ap. J. 282, 671.
\item{}Chernicaro, J., Castets, A., Duvert, G., and Guilliteau, S. 1984,
    Astron. Astrophys. 139, L13.
\item{}Clemens, D.P. and Barvainis, R. 1988,     Ap. J. Suppl. 68, 257 (CB).
\item{}Cohen, M. and Kuhi, L.V. 1979,     Ap. J. Suppl. 41, 743.
\item{}De Lucia, F., and Gordy, W. 1969,     Phys. Rev. 187, 58.
\item{}Emerson, J.P. 1987 in Star Forming Regions, Proc. IAU Symp. 128,
eds. M. Peimbert and J. Jugaku (Dordrecht, Kluwer), p. 19.
\item{}Finkenzeller, U., and Mundt, R. 1984, Astron. Astrophys. Suppl. Ser.
55, 109.
\item{}Frerking, M.A., and Langer, W.D. 1982,     Ap. J. 256, 523.
\item{}G\"usten, R., and Marcaide, J.M. 1986,     Astron. Astrophys. 164, 342.
\item{}Harju, J. 1989,     Astron. Astrophys. 219, 293.
\item{} Hartquist, T.W., Rawlings, J.M.C., Williams, D.A., and Dalgarno, A.
1993, Q.J.R. astr. Soc., 34, 213.
\item{}Harvey, P.M. 1984,     Publ. Astron. Soc. Pac. 96, 297.
\item{}Herbig, G.H. 1960,     Ap. J. Suppl. 4, 337.
\item{}Heyer, M.H., Snell, R.L., Goldsmith, P.E., and Myers, P.C. 1987,
    Ap. J. 321, 370.
\item{}Ho, P.T.P., Moran, J.M., and Rodriguez L.F. 1982,     Ap. J. 262, 619.
\item{}Huntress, W.T., Jr., and Anicich, V.G. 1976,     Ap. J. 208, 237
\item{}Irvine, W.M., Schloerb, F.P., Hjalmarson, \AA. and Herbst, E.
1985, in     Protostars and Planets II, eds. D.C. Black and M.S. Matthews
(Tucson: The University of Arizona Press), p. 579.
\item{}Irvine, W.M., Goldsmith, P.F., and Hjalmarson, \AA. 1987, in
Interstellar Processes, eds. D.J. Hollenbach and H.A. Thronson, Jr.
(Dordrecht: Reidal), p. 561.
\item{}Kogure, T., Yoshida, S., Nakano, T., Tatematsu, T., Jun, X., and
Lanping, X. 1988,     Vistas in Astron. 31, 473.
\item{}Koo, B.C. 1989,     Ap. J. 337, 318.
\item{}Lada, C.J. 1985, Ann. Rev. Astron. Astrophys. 23, 267.
\item{}Lada, C.J., Blitz, L., Reid, M.J., and Moran, J.M. 1981,
Astrophys. J. 243, 769.
\item{}Leung, C.M. 1985 in     Protostars and Planets II, eds. D.C. Black
and M.S. Matthews (Tucson: The University of Arizona Press), p. 104.
\item{}Levreault, R.M. 1988,     Ap. J. Suppl. 67, 283.
\item{}Loren, R.B. 1977,     Ap. J. 218, 716.
\item{}Loren, R.B. 1981,     Astron. J. 86, 69.
\item{}Loren, R.B., Vanden Bout, P.A., and Davies, J.H. 1973,     Ap.
J. 185, 267.
\item{}Lynds, B.T. 1962,     Ap. J. Suppl. 7, 1.
\item{} Myers, P.C. 1985, in Protostars and Planets, eds. D.C. Black
and Matthews (Tucson: The University Press) p. 81.
\item{}Nekel, T., Chini, R., G\"usten, R., and Wink, J.E. 1985, Astron.
Astrophys. 153, 253.
\item{}Parker, N.D., Padman, R., Scott, P.F., and Hills, R.E. 1988, Mon.
Not. R. astr. Soc. 234, 67.
\item{}Pearson, E.F., Creswell, R.A., Winnewisser, M., and Winnewisser, G.
1976,     Z. Naturforsch. 31a, 1394.
\item{}Poetzel, R., Mundt, R., and Ray, T.P. 1992,     Astron. Astrophys.
 262, 229.
\item{}Rodriguez, L.F., Haschick, A.D., Torrelles, J.M., and Myers, P.C. 1987,
    Astron. Astrophys. 186, 319.
\item{}Rodriguez, L.F., Moran, J.M., Ho, P.T.P., and Gottlieb, E.W., 1980,
    Astrophys. J. L35, 845.
\item{}Scappini, F., Caselli, P., and Palumbo, G.G.C. 1991,     Mon. Not.
R. astr. Soc. 249, 763.
\item{}Strom, S.E., Strom, K.M., Yost, J., Carrasco, L., and Grasdalen, G.L.
1972,     Ap. J. 173, 353.
\item{}Sugitani, K., Fukui, Y., Mizuno, A., and Ohashi, N. 1989,     Ap.
J. (Letters) 342, L87.
\item{}Tereby, S., Vogel, S.N., and Myers, P.C. 1989,     Ap. J. 340, 472.
\item{}Walmsley, C.M. 1993,     J. Chem. Soc. Faraday Trans. 89, 2119.
\item{}Wootten, A., Loren, R.B., and Snell, R.L. 1982,     Ap. J.
 255, 160.
\item{}Xie, T., and Goldsmith, P.F. 1990, Ap. J. 359, 378.
\item{}Yoshida, S., Kogure, T., Nakano, M., Tatematsu, K., and
Wiramihardja, S.D. 1991,     Publ. Astron. Soc. Japan 43, 363.
\item{} Yun, J.L., and Clemens, D.P. 1992, Ap. J. 385, L21 (YC).
\end{description}
\newpage
\begin{center} \section*{FIGURE CAPTIONS} \end{center} \vspace*{1cm}
\begin{description}
\item[Figs. 1 to 10] Line profiles of the J=1$\rightarrow$0
transitions of HCO$^+$, H$^{13}$CO$^+$, HCN and H$^{13}$CN in a number of
sources.
The telescope pointing positions correspond to
the coordinates given in Table 1.
\item[Fig. 11] Contour maps of the integrated line intensity
distribution of HCO$^+$
(solid line) and of HCN (dashed line) in CB34. Levels are
\(\frac{1}{\eta_{mb}} \int T_{A}^{*} dv \) = 1,2,3,4,5 KKms$^{-1}$.
\item[Fig. 12] Contour maps of the integrated line intensity
distribution of HCO$^+$
(solid line) and of HCN (dashed line)
in LkH$\alpha$234. Levels are
\(\frac{1}{\eta_{mb}} \int T_{A}^{*} dv \) = 3,5,7,9,11 KKms$^{-1}$.
\end{description}
\newpage
\begin{tabular}{llcrclcrrr}
\multicolumn{10}{c}{TABLE 1} \\ \\
\multicolumn{10}{c}{\small LIST OF THE SOURCES AND OF THEIR PHYSICAL
PROPERTIES} \\ \\
\hline\hline \\
Source && $\alpha_{1950}$  && \multicolumn{1}{c}{   }& &
$\delta_{1950}$ & & \multicolumn{1}{c}{$v^{a}_{LSR}$} & Comments \\
name &&   (hms)     & &\multicolumn{1}{c}{    }& & ($^o$ ' ") & &
(kms$^{-1}$)& \\ \\
\hline \\
\multicolumn{9}{c}{Bok globules} \\ \\
LBN594$^1$ & 00 & 25 & 59.0 && 56 & 25 & 32 & -38.3 & H$_{2}$O maser$^{b}$\\
LBN613$^1$ & 00 & 46 & 34.0 && 50 & 28 & 25 & -12.5 & \\
L1534$^2$ & 04 & 36 & 31.6 && 25 & 35 & 56 & 6.1 & \\
CB34$^1$ & 05 & 44 & 03.0 && 20 & 59 & 07 & 0.7 & \\
L810$^3$ & 19 & 43 & 21.0 && 27 & 43 & 37 & 15.8 & H$_{2}$O maser$^{c}$\\
L797$^1$ & 20 & 03 & 44.0 && 23 & 17 & 54 & 12.6 & \\
L1262$^1$ & 23 & 23 & 48.0 && 74 & 01 & 07 & 4.0 & \\ \\
\multicolumn{9}{c}{Herbig Ae/Be stars} \\ \\
LkH$\alpha$198$^4$ & 00 & 08 & 44.0 && 58 & 33 & 06 & -0.7 & \\
RRTau$^5$ & 05 & 36 & 23.3 && 26 & 20 & 56 & -5.4 & \\
HD250550$^4$ & 05 & 59 & 06.5 && 16 & 30 & 58 & 2.1 & \\
V645Cyg$^4$ & 21 & 38 & 10.6 && 50 & 00 & 43 & -44.6 & H$_{2}$O maser$^{d}$\\
LkH$\alpha$234$^4$ & 21 & 41 & 57.0 && 65 & 53 & 09 & -7.8 &
H$_{2}$O maser$^{e}$\\
BD463471$^4$ & 21 & 50 & 38.5 && 46 & 59 & 34 & 7.0 & \\
LkH$\alpha$233$^6$ & 22 & 32 & 28.2 && 40 & 24 & 33 & 0.1 & \\
MWC1080$^4$ & 23 & 15 & 14.9 && 60 & 34 & 19 & -29.1 & \\ \\ \hline \\
\end{tabular}

REFERENCES.- (1) Clemens and Barvainis 1988; (2) Meyer {\it et al.} 1987; \\
\hspace*{1cm} (3) Nekel {\it et al.} 1985; (4) Lada 1985;
(5) Loren 1981; (6) Levreault 1988. \\

$^a$ From CO millimeter-wave measurements
 in the clouds (CB) and in the star \\
\hspace*{1cm} environments (Cant\`o {\it et al.} 1984). \\
\hspace*{0.7cm}$^b$ Scappini {\it et al.}, 1991.  \\
\hspace*{0.7cm}$^c$ Neckel {\it et al.}, 1985. \\
\hspace*{0.7cm}$^d$ Lada {\it et al.}, 1981. \\
\hspace*{0.7cm}$^e$ Rodriguez {\it et al.}, 1980. \\
\newpage
\begin{center}
TABLE 2 \\              \vspace*{0.5cm}
{\small ANTENNA TEMPERATURE, FWHM LINEWIDTH, AND LSR VELOCITY} \\
{\small OF THE PEAK EMISSION}  \\
\vspace*{0.3cm}
\begin{tabular}{llcrc} \hline \hline \\
Source &  \multicolumn{1}{c}{T$_{A}^{*}$} &  \multicolumn{1}{c}{$\Delta v$}
&  \multicolumn{1}{c}{$v_{LSR}$} & C-isotope \\
name & \multicolumn{1}{c}{(K)} & \multicolumn{1}{c}{(kms$^{-1}$)}
& \multicolumn{1}{c}{(kms$^{-1}$)} &  \\ \hline \\
\multicolumn{5}{c}{HCO$^{+}$} \\ \\
LBN594         & 0.48(5)$^{a}$ & 1.9 & -39.4 & 12\\
               & $<$0.05         &     &       & 13   \\
LBN613         & 0.63(3) & 1.1 & -12.4 & 12\\
L1534          & 0.30(4) & 2.0 & 6.7 & 12\\
CB34           & 0.91(4) & 2.0 & 0.5 & 12  \\
               & $<$0.07 &     &     & 13 \\
L810           & 1.75(6) & 2.0 & 15.9 & 12\\
               & 0.20(4) &     &      & 13 \\
L797           & 0.13(2) & 2.2 & 12.2 & 12\\
L1262          & 0.47(4) & 1.6 & 4.4  & 12\\ \\
LkH$\alpha$198 & 0.99(3) & 1.3 & 0.0  & 12\\
               & 0.27(4) &     &      & 13 \\
RRTau          & $<$0.09 &     &      & 12 \\
HD250550       & $<$0.11 &     &      & 12\\
V645Cyg        & 1.63(5) & 3.1 & -43.9& 12 \\
               & 0.28(2) &     &      & 13 \\
LkH$\alpha$234 & 2.15(6) & 2.5 & -10.3& 12  \\
               & 0.16(2) &     &      & 13 \\
BD463471       & 0.22(3) & 1.7 & 6.8  & 12\\
LkH$\alpha$233 & 0.20(5) & 1.4 & -0.1 & 12\\
MWC1080        & 0.30(3) & 2.4 & -30.1& 12 \\
               & $<$0.18 &     &      & 13 \\ \\
\end{tabular}
\end{center}
\newpage
\begin{center}
TABLE 2 (continued) \end{center}
\begin{center}
\begin{tabular}{llrrc}
 \multicolumn{5}{c}{HCN} \\ \\
LBN594         & 0.79(10)& 2.3 & -40.2 & 12\\
LBN613         & 0.25(6) & 1.3 & -12.4 & 12\\
L1534          & $<$0.28 &     &       & 12\\
CB34           & 0.83(5) & 1.8 &  0.5  & 12\\
               & $<$0.10 &     &       & 13 \\
L810           & 1.35(14)& 1.4 &  16.0 & 12\\
L797           & $<$0.11 &     &       & 12\\
L1262          & 0.55(4) & 2.1 &  4.1  & 12\\ \\
LkH$\alpha$198 & 0.60(9) & 1.3 &  0.1  & 12\\
RRTau          & $<$0.89 &     &       & 12 \\
HD250550       & $<$0.08 &     &       & 12\\
V645Cyg        & 1.70(7) & 2.7 & -43.9 & 12\\
               & 0.09(1) &     &       & 13 \\
LkH$\alpha$234 & 1.56(6) & 2.4 & -10.2 & 12\\
               & $<$0.07 &     &       & 13 \\
BD463471       & $<$0.10 &     &       & 12\\
LkH$\alpha$233 & 0.23(7) & 1.7 & 0.0   & 12\\
MWC1080        & 0.43(4) & 2.4 & -30.5 & 12 \\ \\ \hline
\end{tabular}
\vspace*{0.5cm}
\end{center}
$^a$Standard errors are in units of the last digit. \\
\newpage
\begin{tabular}{lrrrrrr}
\multicolumn{7}{c}{TABLE 3} \\ \\
\multicolumn{7}{c}{\small INTEGRATED LINE INTENSITIES
 $I(X)^a$
 FOR THE OBSERVED SPECIES.} \\
\multicolumn{7}{c}{\small RATIOS (R) OF THE INTENSITIES FOR THE $^{12}$C
AND $^{13}$C ISOTOPOMERS} \\ \\
\hline\hline \\
Source&\multicolumn{1}{c}{I(HCO$^+$)}&\multicolumn{1}{c}{I(H$^{13}$CO$^+$)}
& \multicolumn{1}{c}{R} & \multicolumn{1}{c}{I(HCN)} &
\multicolumn{1}{c}{I(H$^{13}$CN)} & \multicolumn{1}{c}{R} \\
name  &\multicolumn{1}{c}{(Kkms$^{-1}$)}&\multicolumn{1}{c}{(Kkms$^{-1}$)}
&  & \multicolumn{1}{c}{(Kkms$^{-1}$)} & \multicolumn{1}{c}{(Kkms$^{-1}$)}
& \\ \\
\hline \\
LBN594 & 2.23(23) & $<$0.51 & $>$4.4 & 4.50(16) & & \\
LBN613 & 1.88(17) & & & 0.79(18) & & \\
L1534 & 1.47(21) & & & $<$2.1 & & \\
CB34 & 4.83(21) & $<$0.73 & $>$6.6 & 3.70(21) & $<$0.61 & $>$6.8 \\
L810 & 8.86(30) & 0.74(18) & 12.0(29) & 4.79 & & \\
L797 & 0.68(20) & & & $<$0.55 & & \\
L1262 & 2.22(23) & & & 2.98(23) & & \\ \\
LkH$\alpha$198 & 3.18(17) & 0.68(12) & 4.7(9) & 1.80(30) & & \\
RRTau & $<$0.66 & & $<$0.65 & & \\
HD250550 & $<$0.83 & & &$<$0.74 & & \\
V645Cyg & 14.28(55) & 1.69(18) & 8.4(10) & 11.46(46) & 0.52(8) & 22.0(35) \\
LkH$\alpha$234 & 12.66(30) & 0.64(12) & 19.8(35) & 9.46(24) & $<$0.45
& $>$21 \\
BD463471 & 1.04(18) & & & $<$0.87 & & \\
LkH$\alpha$233 & 0.70(16) & & & 0.98(32) & & \\
MWC1080 & 1.78(11) & $<$1.07 & $>$1.7 & 2.55(11) & & \\ \\
\hline
\end{tabular}

\vspace*{0.5cm}
$^{^a}$$ I(X)=\frac{1}{\eta_{mb}} \int T_{A}^{*} dv $ for $X$ = HCO$^+$,
H$^{13}$CO$^+$, HCN, and H$^{13}$CN, respectively.
\newpage
\begin{center}
TABLE 4 \\ \vspace*{0.5cm}
{\small OPTICAL DEPTH AND EXCITATION TEMPERATURE FOR THE SOURCES} \\
{\small FOR WHICH DATA EXIST RELATIVE TO THE $^{13}$C ISOTOPOMERS}  \\
\vspace*{0.5cm}
\begin{tabular}{lrrl} \hline \hline \\
Source &  \multicolumn{1}{c}{$\tau_{13}^{a}$} &
\multicolumn{1}{c}{$\tau_{12}^{b}$} &  \multicolumn{1}{c}{T$_{ex}$}  \\
name &  &  & \multicolumn{1}{c}{(K)}  \\ \\
\hline \\
\multicolumn{4}{c}{HCO$^{+}$} \\ \\
LBN594 & $<$0.26 & $<$22.9 & \\
CB34 & $<$0.16 & $<$14.6 & \\
L810 & 0.09(2) & 7.7(17) & 7.0 \\ \\
LkH$\alpha$198 & 0.24(4) & 21.4(40) & 5.1 \\
V645Cyg & 0.13(2) & 11.3(17) & 6.7 \\
LkH$\alpha$234 & 0.05(1) & 4.4(9)   & 7.8 \\
MWC1080        & $<$0.89 & $<$79 & \\ \\
\multicolumn{4}{c}{HCN} \\ \\
CB34 & $<$0.16 & $<$2.2 \\ \\
V645Cyg & 0.05(1) & 4.1(8) & 6.9 \\
LkH$\alpha$234 & $<$0.05 & $<$4.3 &  \\ \hline \\
\end{tabular}
\end{center}

\noindent
$^a$$\tau_{13}$ refers to the $^{13}$C isotopomer of H$^{13}$CO$^+$
or H$^{13}$CN. \\
$^{b}$$\tau_{12}$ refers to the $^{12}$C isotopomer of HCO$^{+}$ or HCN.
\newpage
\begin{tabular}{lrrrr}
\multicolumn{5}{c}{TABLE 5} \\ \\
\multicolumn{5}{c}{\small COLUMN DENSITIES OF HCO$^+$, H$^{13}$CO$^+$,
HCN, AND H$^{13}$CN } \\ \\
\hline\hline \\
 &\multicolumn{1}{c}{N[HCO$^+$]}& \multicolumn{1}{c}{N[H$^{13}$CO$^+$]}
 & \multicolumn{1}{c}{N[HCN]} & \multicolumn{1}{c}{N[H$^{13}$CN]} \\
 &\multicolumn{1}{c}{(cm$^{-2}$)$\times$10$^{+12}$}
 &\multicolumn{1}{c}{(cm$^{-2}$)$\times$10$^{+12}$}
 &\multicolumn{1}{c}{(cm$^{-2}$)$\times$10$^{+12}$}
 &\multicolumn{1}{c}{(cm$^{-2}$)$\times$10$^{+12}$}\\ \\
\hline \\
LBN594          & 2.16(22)& $<$0.51   &  7.96(24) &       \\
LBN613          & 1.82(16)&           & 1.40(32)  &       \\
L1534           & 1.43(20)&           & $<$3.72   &       \\
CB34            & 4.68(20)& $<$0.71   & 6.55(37)  & $<$1.15 \\
L810            & 66.2(22)& 0.74(18)  & 65.3(70)$^a$ &    \\
L797            & 0.66(19)&           & $<$1.50   &       \\
L1262           & 2.15(22)&           & 5.27(41)  &       \\ \\
LkH$\alpha$198  & 66.0(35)& 0.69(12)  & 68.2(11)$^a$  &   \\
RRTau           & $<$0.64 &           & $<$1.15   &       \\
HD250550        & $<$0.80 &           & $<$1.31   &      \\
V645Cyg         & 156.5(60)& 1.71(18) & 83.2(32)  & 0.98(15) \\
LkH$\alpha$234  & 54.0(13)& 0.65(12)  & 73.7(19)$^a$  & $<$0.85   \\
BD463471        & 1.01(17)&           & $<$1.54   &          \\
LkH$\alpha$233  & 0.68(15)&           & 1.73(57)  &       \\
MWC1080         & 1.73(11)& $<$1.08   & 4.51      &       \\ \\
\hline \\
\end{tabular}

\hspace*{0.7cm}$^a$Assuming $\tau_{12}$(HCN) $\simeq$ $\tau_{12}$(HCO$^+$).
\newpage
\begin{center}
TABLE 6 \\ \vspace*{0.5cm}
{\small THE HCO$^+$/HCN COLUMN DENSITY RATIO FOR DIFFERENT REGIONS}\\
{\small TOGETHER WITH EXCITATION TEMPERATURE, DENSITY  AND} \\
{\small LINEWIDTH INFORMATION} \\
\vspace*{0.5cm}
\begin{tabular}{lccccc}  \hline \hline \\ \\
 \multicolumn{1}{c}{Source}& \multicolumn{1}{c}{HCO$^+$/HCN} &
 \multicolumn{1}{c}{$T(K)$} & \multicolumn{1}{c}{$n$(cm$^{-3}$)} &
 \multicolumn{1}{c}{$\Delta v$(kms$^{-1}$)$^{a}$}
 & \multicolumn{1}{c}{Ref.} \\ \\
\hline  \\ \\
TMC-1           & 0.4     & 10-20     & 10$^3$-10$^5$ & 0.2-0.9 & 1  \\ \\
L134N(L183)     & 2.0     & 10-20     & 10$^3$-10$^5$ & 0.2-0.9 & 1  \\ \\
Orion ridge     & 0.15    & 50-100    & 10$^4$-10$^6$ & 2.5-4 & 1      \\ \\
Orion plateau   & 0.03    & 90-150    & 10$^6$-10$^7$ & 20 & 1 \\ \\
Bok globules    & 0.3-1.3 & 10        & 10$^3$-10$^4$ & 1.1-2.2 & 2 \\ \\
Herbig Ae/Be stars & 0.4-1.9 & 10     &               & 1.3-3.1 & 2 \\ \\
\hline \\ \\
\end{tabular}
\end{center}

\noindent
$^a$FWHM linewidth. \\
$^1$Irvine {\it et al.}, 1985; Irvine {\it et al.}, 1987. \\
$^2$Present work. The density of the Bok globules is taken
from Leung,
1985.
\end{document}